# Establishing relationship between an uniaxial pressure and an electric field shifting the Curie temperature in BaTiO$_3$ crystals

E. Dul'kin*, M. Roth

*Department of Applied Physics, The Hebrew University of Jerusalem, Jerusalem 91904, Israel*

Shifting the Curie temperature in dependence on both uniaxial pressure and electric field in BaTiO$_3$ crystals was studied based on literature data. It was shown that both these dependences perfectly coincide when adjusting the scale. Based on coincidence of these dependencies a relationship between both an uniaxial pressure and an electric field when shifting the Curie temperature was established.

*to whom correspondence should be addressed: evgeniy.dulkin@mail.huji.ac.il

Barium titanate, $BaTiO_3$, is a well-known ferroelectric, has been discovered in 1945 and since is extensively studied. It is well documented that upon heating $BaTiO_3$ undergoes a structural transition from ferroelectric to paraelectric phase pointed out by sharp peak of dielectric permittivity at the Curie temperature, $T_c$. An effect of both electric field, $E$, and mechanical pressure, $p$, on $T_c$ shift is studied as well.

Shifting the $T_c$ on 8.5ºC, detected from the hysteresis loops of $BaTiO_3$ crystals grown by the Remeika method, was found to be a linear as $E$ enhances up to 6 kV/cm at the fixed temperatures up to 116ºC [1]. Meanwhile, shifting the $T_c$ on 15ºC, detected from the birefringence of $BaTiO_3$ crystals grown by the Remeika method, was found to be a nonlinear as $E$ enhances up to 12 kV/cm and a linear as $E$ weakens down to 0 kV/cm at the fixed temperatures up to 136ºC [2]. However, shifting the $T_c$ on 3ºC, detected by the acoustic emission of $BaTiO_3$ crystals grown by the melt-grown method, was found to be a linear upon heating at the fixed fields up to 2 kV/cm [3]. Recently, shifting the $T_c$ on 8.5ºC, detected from the electrocaloric effect of $BaTiO_3$ crystals grown by the melt-grown method, was found to be a nonlinear upon heating at the fixed fields up to 10 kV/cm [4]. While the dependencies between $T_c$ and $E$ in Refs. [1,3] coincide well, the dependencies between $T_c$ and $E$ in Refs. [2,4] are not consistent.

Shifting the $T_c$ on 3ºC, detected by the dielectric permittivity of $BaTiO_3$ crystals grown by the melt-grown method, upon heating up to 410ºC at the fixed uniaxial pressures up 1000 bar was measured previously [5]. Shifting the $T_c$ in dependence on $p$ was approximated to be a linear, but one can clearly see that it is a very rough approximation. In fact the $T_c(p)$ dependence is a nonlinear and visibly trends to saturation as the $p$ enhances. Also a relationship between $p$ and $E$, p/E one can calculate to be 23.8 bar·cm/kV at room temperature, not at $T_c$, as it might be expected.

The goal of the present paper to estimate the relationship between both $p$ and $E$ within the $T_c$ shifting region based on comparison the data of above cited works.

In this paper a consideration is devoted to comparison the data of BaTiO$_3$ crystals, used in the Refs. [4,5], because they were grown by the same melt-grown method, and exhibit the same $T_c \approx 407$ K, and studied at the same conditions: upon heating under fixed pressure and field up to their higher values, and, thus, can be compared truly.

Fig. 1. presents the $T_c$ shifting in dependence on both $p$ and $E$ reconstructed from the corresponding data of Refs. [5,4], respectively. Accurate reconstructed the $T_c$ shifting in dependence on $p$ is indeed a nonlinear in contrast to that declared in Ref. [5]. One can see that both these dependencies perfectly coincide when adjusting the scale. Such the perfect coincidence unambiguously proves that both $p$ and $E$ shift the $T_c$ in the same manner.

Both these $T_c(p)$ and $T_c(E)$ dependencies are approximated the following equations:

for uniaxial pressure

$$T_c^p = 407 + 3.2 \cdot 10^{-3} p - 7.3 \cdot 10^{-7} p^2$$

for electric field

$$T_c^E = 407 + E - 2.29 \cdot 10^{-2} E^2$$

From these equations one can establish the relationship between both $p$ and $E$ values at the same $T_c$. For example, to shift the $T_c$ on 2 K, i.e. up to 409 K one need apply the or equivalent $p = 750$ bar or the equivalent $E = 2$ kV/cm, and, consequently, the relationship between both $p$ and $E$, $p/E$ is about 375 bar·cm/kV.

In summary, we have compared the Curie temperature shifting in dependence on both uniaxial pressure and electric field based on literature data and found their acting proportionally the same. Based on this proportionality we have established the relationship is equal to be 375 bar·cm/kV between both uniaxial pressure and electric field when shifting the Curie temperature.

**Figure Caption**

Fig. 1. A complex plot of $T_c$ shifting in dependence on an uniaxial pressure, $p$, (filled squares, thick dash) and on an electric field, $E$, (filled circles, thin dash).

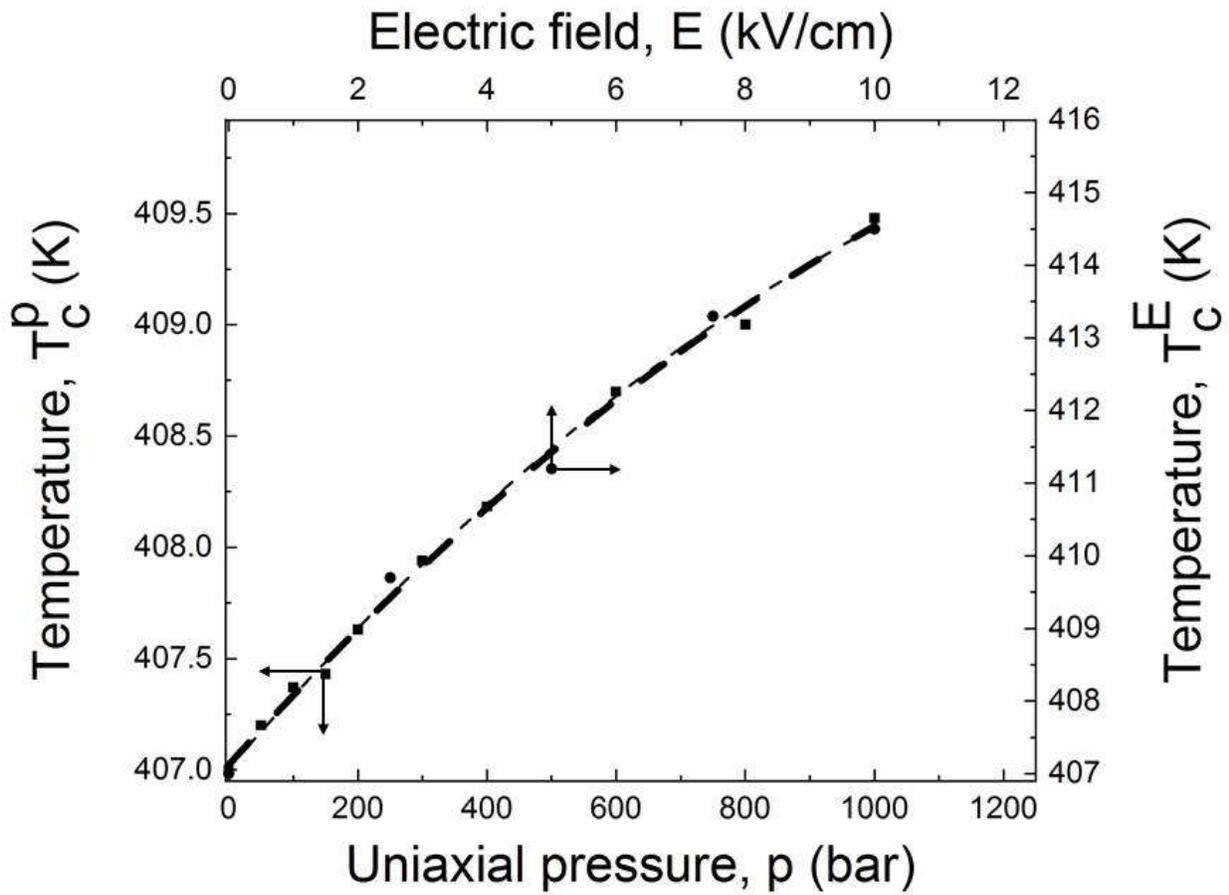